\documentstyle[epsfig]{elsart}
\newbox\mybox

\newcommand\fverb{\setbox\mybox=\hbox\bgroup\verb}
\newcommand\fverbdo{\egroup\medskip\noindent\fbox{\unhbox\mybox}\ }
\newcommand\fverbit{\egroup\item[\fbox{\unhbox\mybox}]}

\newcommand\init[1]{\setbox\mybox=\hbox{{\beeg #1}~}%
                   \noindent\global\hangindent=\wd\mybox\global\hangafter-2
                   \sc\smash{\llap {\lower 13.2pt \box\mybox}}}
\def\v1{\vspace{1cm}}
\def\be{\begin{equation}}
\def\ee{\end{equation}}
\def\bc{\begin{center}}
\def\ec{\end{center}}
\def\vh{\varphi}

\newcommand{\bea}{\begin{eqnarray}}
\newcommand{\eea}{\end{eqnarray}}

\begin{document}
\begin{frontmatter}
\title{CMBR anisotropy: theoretical approaches}
\author[dub]{B.M. Barbashov,}
\author[dub]{V.N. Pervushin,}
\author[itep]{A.F. Zakharov,}
\author[dub]{V.A. Zinchuk}
\address[dub]{Bogoliubov Laboratory for Theoretical Physics, JINR, 141980 Dubna,
Russia}
\address[itep]{Institute of Theoretical and Experimental Physics, B. Cheremushkinskaya, 25,
117259, Moscow, Russia}
\begin{abstract}

 The version of the cosmological perturbation theory
 based on exact resolution of energy constraint
 is developed in accordance with the diffeomorphisms of general relativity
 in the Dirac Hamiltonian approach.
  Such exact resolution gives one
 a possibility to fulfil the Hamiltonian reduction and
 to explain the ``CMBR primordial power spectrum'' and other topical
problems of modern cosmology by quantization of the energy
constraint and quantum cosmological origin of matter.
\end{abstract}
\begin{keyword}
General Relativity and Gravitation, Quantum field theory,
Observational Cosmology
\\[2mm]
\end{keyword}
\end{frontmatter}
%

\section{Introduction}

 One of the basic tools applied for analysis of modern
observational data including Cosmological Microwave Background
Radiation (CMBR) is the cosmological perturbation theory in
general relativity (GR)  \cite{lif,bard}. The main role in this
perturbation theory plays the  separation of the cosmological
scale factor by the transformation
$g_{\mu\nu}=a^2(x^0)\overline{g}_{\mu\nu}$. The similar separation
\cite{prd,ps1,pp,bpp} of the  homogeneous scale factor $a$ is also
fulfilled in the Hamiltonian approach \cite{dir,ADM} to GR, where
$a$ is an invariant evolution parameter in accordance with the
Hamiltonian diffeomorphism subgroup $x^0 \to
\widetilde{x}^0=\widetilde{x}^0(x^0)$ \cite{vlad}  meaning in fact
that the coordinate ``time'' $x^0$ is not observable.
 In this case, in order to keep
 the number of
 variables of GR,
  such the separation of the scale factor $a$ should
  be accompanied by the constraint
   that removes a
  similar homogeneous variable from the rest spatial metric
  determinant.
 Just the separation of homogeneous scale factor from
 metrics (in contrast with the Dirac Hamiltonian approach  \cite{dir}, where
 the scale factor $a$ is not taken into account), on the
 one hand, and the keeping the  number of variables of GR
 (in contrast with the  standard perturbation theory
 \cite{lif,bard}, where
 the scale factor $a$ is  taken into account twofold), on other
 hand,
 gives rise to exact resolution of the energy constraint \cite{pp}
 $\overline{g}^{00}\sqrt{-\overline{g}}/\int d^3x\overline{g}^{00}\sqrt{-\overline{g}}
 =
 \sqrt{T_0^0}/\int d^3x\sqrt{T_0^0}$, where $T_0^0$ is the
 energy-momentum tensor component  in terms of the Lichnerowicz
 conformal variables \cite{L,Y}.
 In this paper the   perturbation theory based on this
exact resolution of the energy constraint in GR is developed and
compared with the standard  cosmological perturbation theory
\cite{lif,bard}. The content of the paper
 is the following.
 In Section 2,
  the Hamiltonian approach to GR with
 the separated cosmological scale  is considered.
 In Section 3,
  the Hamiltonian   perturbation theory is formulated in a finite space-time
  as an alternative of the cosmological perturbation theory.
 Two possible
 descriptions of CMBR
  are compared in  Section 4.

\section{\label{sf}The diffeomorphism-invariant Hamiltonian approach to GR}

 Let us  consider the Hilbert - Einstein action of GR
 \be\label{gr}S[\vh_0]=\int d^4x\sqrt{-g}\left[-\frac{\vh_0^2}{6}R(g)
 +{\cal L}_{(\rm M)}(\vh_0|g,f)\right]
 \ee
 in a space-time with the interval $ds^2=g_{\mu\nu}dx^\mu
 dx^\nu$, where $\varphi_0=\sqrt{3/{8\pi G_0}}=\sqrt{3M^2_{\rm
 Planck}/{8\pi}}$ is
 the Newton coupling constant that scales all masses.
 The Hamiltonian approach to GR is formulated
  in a frame of reference  given by a geometric interval
 $
 g_{\mu\nu}dx^\mu dx^\nu\equiv\omega_{(0)}\omega_{(0)}-
 \omega_{(1)}\omega_{(1)}-\omega_{(2)}\omega_{(2)}-\omega_{(3)}\omega_{(3)}
 $, where
   $\omega_{(\alpha)}$ are  linear differential forms
  \cite{fock29}  in
 terms of the Dirac variables \cite{dir}
\be \label{adm}
 \omega_{(0)}=\psi^6N_{\rm d}dx^0,~~~~~~~~~~~
 \omega_{(b)}=\psi^2 {\bf e}_{(b)i}(dx^i+N^i dx^0);
 \ee
 here $\psi$ is the spatial metrics determinant variable,
  triads ${\bf e_{(a)i}}$ form the spatial metrics with $\det |{\bf
 e}|=1$, $N_{\rm d}$ is the lapse function, and $N^i$ is the shift vector.

 The forms (\ref{adm}) are invariant with respect to
 the kinemetric general coordinate transformations
 $\widetilde{x}^0=\widetilde{x}^0(x^0),
 \widetilde{x}^i=\widetilde{x}^i(x^0,x^l)$ \cite{vlad}.
 This  group of diffeomorphisms of the  frame (\ref{adm})
 means that the coordinate time is
 not observable.
 One of the main problems of the Hamiltonian approach to GR is to
 pick out a diffeomorphism(d)-invariant global variable which can be
{\it``evolution parameter''}.
 There is a set of arguments  \cite{prd,ps1,pp,bpp,Y} to   identify  this
 ``evolution parameter'' in GR with the  cosmological
 scale factor  $a(x_0)$ introduced
 by the scale transformation:
 $F^{(n)}=a^n(x_0) \overline{F}^{(n)}$, where $(n)$ is the conformal weight
 of any field, including
 metric components, in particular the  lapse
 function ${N}_d=a^{-2}\overline{N}_d$ and
 $\psi^2=a\overline{\psi}^2\!.$
 The logarithm form $\log{\psi^2}=F$ of this equation can be presented
as a sum $F=\langle F\rangle +[F-\langle F\rangle]$, where
$\left\langle{ F}\right\rangle \equiv \int d^3x
 F/V_0$ identified
with $\log a$ ($V_0=\int d^3x  < \infty$ is a finite
 volume) is the  spatial volume ``averaging'' and $[F-\langle F\rangle]=\overline{F}$ is the
orthogonal operation of
 ``deviation''
  distinguished by the identity\footnote{Just this identity
  is the main difference of our approach to GR
  from the Lifshitz perturbation theory \cite{lif,bard}.}
 \be\label{non1}
 \int d^3x \log\overline{\psi}^2(x^0,x^i)=\int d^3x \left[\log{\psi^2}
 -\left\langle{ \log{\psi^2}}\right\rangle\right]\equiv 0.
 \ee
  The transformation of a curvature
 $
 \sqrt{-g}R(g)=a^2\sqrt{-\overline{g}}R(\overline{g})-6a
 \partial_0\left[{\partial_0a}\sqrt{-\overline{g}}~\overline{g}^{00}\right]$
  converts action (\ref{gr}) into
 \be\label{1gr}
 S[\vh_0]=\overline{S}[\vh]-
 \int dx^0 (\partial_0\vh)^2\int \frac{d^3x}{\overline{N}_d}
 \equiv \int dx^0 L\equiv \int d^4x {\cal L},
 \ee
 here $\overline{S}[\varphi]$
  is the action (\ref{gr})  in
 terms of metrics $\overline{g}$, where  $\vh_0$ is replaced by
 the running scale $\vh(x^0)=\vh_0a(x^0)$ of all masses  of the
 matter fields.
 The
 energy constraint ${\delta S[\vh_0]}/{\delta  \overline{N_d}}=0$ takes the
 algebraic form \cite{pp}:
 \be\label{nph}
 ~~~~~~~~~~~~~~~~~~~~~~~~-
 \frac{\delta \overline{S}[\vh]}{\delta  \overline{N_d}}
 \equiv T^0_0=\frac{(\partial_0\varphi)^2}{\overline{N_d}^2}
 \ee
 where $T^0_0$ is the local energy density. This equation has exact solution:
\be\label{13ec}
 \left[\frac{d\varphi}{d\zeta}\right]^2\equiv\vh'^2=
 {\left\langle\sqrt{{T^0_0}}\right\rangle}^2,~~~~~~~~~~{\langle(\overline{N_d})^{-1} \rangle
 \overline{N_d}}=\frac{{\left\langle\sqrt{{T^0_0}}\right\rangle}}{\sqrt{{T^0_0}}},
 \ee
 where
\be\label{13c}
 \zeta(\varphi_0|\varphi)
 \equiv\int dx^0 \langle{ (\overline{N_d})^{-1}}\rangle^{-1}
 =\pm
 \int_{\vh}^{\vh_0}
 \frac{d\widetilde{\vh}}{{\langle \sqrt{T_0^0(\widetilde{\vh})}\rangle}}
 \ee
 is d-invariant time.
This evolution can be treated as
 the analogy of the Hubble law in the d-invariant formulation of exact GR.

 One can construct the Hamiltonian function using the definition of a set of the canonical
 momenta, including
\bea \label{pph}
 P_\vh&=&-\frac{\partial L}{\partial (\partial_0\vh)}
 =2\frac{d\vh}{dx^0}\int \frac{d^3x}{\overline{N_d}}
 \equiv 2V_0 \vh'\\
 \label{gauge}
 p_{\psi}&=&\frac{\partial {\cal L}}{\partial (\partial_0\log\overline{\psi})}\equiv
 -\frac{4\vh^2}{3}\cdot\frac{\partial_l(\overline{\psi}^{6}N^l)-
 \partial_0(\overline{\psi}^{6})}{\overline{\psi}^{6}\overline{N_d}},
 \eea
  if the strong constraint $\langle p_\psi\rangle=0$ is imposed
   to be consistent with the identity
 (\ref{non1}) and to keep the number of variables of GR,
  otherwise the double counting of the zero-Fourier
 harmonics of spatial metric determinant  \cite{lif,bard} does not admit
 resolution of velocities in terms of momenta (that is the obstacle
 for the Hamiltonian approach).
Now, using solution (\ref{13ec}) and definitions (\ref{pph}),
(\ref{gauge}) one can express action in Hamiltonian form in terms
of momenta $P_\vh$ and  $P_{ F}=[{p_{\psi}}, p^i_{{(a)}},p_f]$
 \be\label{hf}
S[\varphi_0]=\int dx^0\left[\int d^3x \left(\sum_F P_F\partial_0
F+C\right)-P_{\varphi}\partial_0\varphi+
\frac{P_{\varphi}^2-E_{\varphi}^2}{4\int dx^3
(\overline{N_d})^{-1}}\right], \ee where the reduced Hamiltonian
function \be\label{hf1} ~~~~~~~~~~~~~~~~~~~~E_\vh=2\int
d^3x\sqrt{{T^0_0}}= 2V_0{\left\langle\sqrt{{T^0_0}}\right\rangle}
 \ee
 can be treated as the ``universe energy'' by analogy with the ``particle energy'' in
 special relativity (SR), and
 ${\cal C}=N^i {T}^0_{i} +C_0p_{\psi}+ C_{(a)}\partial_k{\bf
e}^k_{(a)}$
  is the sum of constraints
  with the Lagrangian multipliers $N^i,C_0,~C_{(a)}$ and the energy--momentum tensor
  components $T^0_i$; these constraints include
   the transversality  $\partial_i {\bf e}^{i}_{(a)}\simeq 0$ and the Dirac
 minimal space-like surface \cite{dir}:
 \be\label{hg}
~~~~~~~~~~~~~~{p_{\psi}}\simeq 0 ~~~~\Rightarrow ~~~~
\partial_j(\overline{\psi}^6{\cal N}^j)=(\overline{\psi}^6)'~~~~~ ({\cal N}^j=N^j\langle
N^{-1}_d\rangle).
 \ee
 One can find
 evolution of all field variables $F(\vh,x^i)$  with respect to
 $\vh$ by the variation of the ``reduced'' action obtained as
   values of the Hamiltonian form of initial action  (\ref{hf}) onto
 the energy constraint $P^2_{\varphi}=E^2_{\varphi}$ \cite{pp}:
 \be\label{2ha2} S[\vh_0]|_{P_\vh=\pm E_\vh} =
 \int\limits_{\vh}^{\vh_0}d\widetilde{\vh} \left\{\int d^3x
 \left[\sum\limits_{  F}P_{  F}\partial_\vh F
 +\bar{\cal C}\mp2\sqrt{T_0^0(\widetilde{\vh})}\right]\right\},
\ee
 where $\bar{\cal C}={\cal
 C}/\partial_0\widetilde{\vh}$.  The reduced Hamiltonian $\sqrt{T_0^0}$ is Hermitian
 as the  minimal surface
 constraint
 (\ref{hg}) removes negative
 contribution of $p_{\psi}$ from energy density.
Thus, the d-invariance   gives us the
 solution of the problem of energy  in GR by the
 Hamiltonian reduction like  solution of the similar problem in
 SR.

  The explicit dependence of $T_0^0$ on $\overline{\psi}$
  was given in
  \cite{L,Y} by extracting the
 Laplace operator $\hat \triangle
 F\equiv\frac{4\varphi^2}{3}\partial_{(b)}\partial_{(b)}F$:
 $T^0_0= \overline{\psi}^{7}\hat \triangle \overline{\psi}+
  \sum_I \overline{\psi}^I\tau_I $, where $I$ runs a set of values
   I=0 (stiff), 4 (radiation), 6 (mass), 8 (curvature), 12
   ($\Lambda$-term)
in the correspondence with a type of matter field contributions.

\section{\label{d-ipt}The diffeomorphism-invariant perturbation theory}

Let's introduce a parametrization of metric
$(\overline{N}_d)^{-1}, \overline{\psi}$ through functions
$\mu,\nu$  \be~~~~~~~~~~ \frac{(\overline{N_d})^{-1}}{\langle
(\overline{N_d})^{-1}\rangle}=1+\nu, ~~~\overline{\psi}=e^{\mu},
 ~~~\langle\mu\rangle\equiv0,~~\langle\nu\rangle\equiv0 \ee
 with the zero spatial ``averaging'' and define partial energy densities
$\tau_I$  as a sum of the ``averaging`` one $\langle
\tau_I\rangle$ and the  ``deviation'' (\ref{non1})
$\overline{\tau}_I=\tau_I-\langle \tau_I\rangle$ that is
orthogonal to the ``averaging`` one $\langle \tau_I\rangle$, so
that the total density takes a form
 \bea\label{2ha4}
 T_0^0 \equiv e^{7\mu}\hat \triangle e^{\mu}+\sum\limits_{I}^{}e^{\mu
I}\tau_I,~~~~~~~\tau_I=\langle \tau_I\rangle+\overline{\tau_I},~~~
\langle \overline{\tau_I}\rangle\equiv0.
 \eea
 The functions $\mu,\nu$  are determined by the
  constraint (\ref{13ec})
   \be \label{14ec} 1+\nu=\frac{\sqrt{{T^0_0}}}
 {\left\langle\sqrt{{T^0_0}}\right\rangle}
 \equiv\frac{\sqrt{e^{7\mu}\hat \triangle e^{\mu}+ \sum\limits_{I}^{}e^{\mu
I}\tau_I}}
 {\left\langle\sqrt{e^{7\mu}\hat \triangle e^{\mu}+
 \sum\limits_{I}^{}e^{\mu I}\tau_I}
 \right\rangle}
 \ee
and the  equation  ${\delta S[\vh]}/{\delta \mu}=-\overline{{\cal
D}}=\langle {\cal D}\rangle-{\cal D}=0$ of spatial metric
variable, where
 \bea \label{4f4}
 \overline{{\cal D}}=\overline{(1+\nu)^{-1}
 \left[\sum\limits_{I}^{}Ie^{\mu I}\tau_I+7e^{7\mu}\hat \triangle  \cdot e^{\mu}\right]+
 e^{\mu}  \hat \triangle  \cdot  [e^{7\mu}(1+\nu)^{-1}]}=0.
 \eea
The Dirac constraint of minimal surface (\ref{hg})
 takes the form
 \be\label{dgauge}
  \partial_{(b)}\left(e^{6\mu}{\cal N}_{(b)}\right)=
 \partial_\zeta(e^{6\mu}).
 \ee
 The d-invariant perturbation theory can be naturally defined as a
  power
 series in ``deviations'' of densities $\overline{\tau}_I$
 defined in the class of functions with the nonzero Fourier harmonics
$\widetilde{\phi}(k)=\int d^3x \overline{\phi}(x)e^{ikx}$
  satisfying constraint $\langle \phi\rangle=0$:
  \bea \label{pt1}
 T_0^0&=&T_0+T_1+T_2+...,\\\label{pt2}
 T_0&=&\langle \tau_{(0)}\rangle,\\\label{pt3}
 T_1&=&\overline{\tau_{(0)}}+(\langle \tau_{(1)}\rangle+\hat \triangle)\cdot\mu,\\\label{pt4}
 T_2&=&\mu\left[\overline{\tau_{(1)}}+(\langle \tau_{(2)}\rangle+14\hat
 \triangle)\cdot\frac{\mu}{2}\right]+\frac12\hat\triangle\cdot(\mu^2),
  \eea
   where $\tau_{(n)}=\sum_I I^n
 \tau_I$. In the first order, Eq. (\ref{14ec})  takes
  a form
  \be\label{ptr1}
 1+\nu=1+\frac{T_1}{2T_0}=1+\frac{\overline{\tau_{(0)}}+(\langle \tau_{(1)}\rangle+\hat
 \triangle)\cdot \mu}{2\langle \tau_{(0)}\rangle}.
 \ee
 The
substitution of this value of $\nu$ into the first order of Eq.
(\ref{4f4}) $\overline{\tau}_{(1)}+(\langle
\tau_{(2)}\rangle+14\hat
 \triangle)\cdot{\mu}-(\langle \tau_{(1)}\rangle+\hat \triangle)\cdot\nu=0$
gives $\nu$ and $\mu$ in the form of sum of the Green functions
$[D\cdot J](x)=\int d^3y D(x-y)J(y)$: \bea\label{2-17}
 {\mu}&=&\frac{1}{14\beta}\left[D_{(+)}\cdot \overline{J_{(+)}}-
 D_{(-)}\cdot \overline{J_{(-)}}\right],\\\label{2-18}
 {\nu}&=&\frac{1}{2\beta}\left[(1+\beta)D_{(+)}\cdot \overline{J_{(+)}}-
 (1-\beta)D_{(-)}\cdot \overline{J_{(-)}}\right],
 \eea
 where $\beta=\sqrt{1+[\langle \tau_{(2)}\rangle-14\langle
\tau_{(1)}\rangle]/(7\langle \tau_{(0)}\rangle)}$,
 \be\label{cur1}
 \overline{J_{(\pm)}}=7(1\pm\beta)\overline{\tau_{(0)}}-\overline{\tau_{(1)}}
 \ee
 are the local currents, $D_{(\pm)}$ are the Green functions satisfying
 equations
 \bea\label{2-19}
 [\pm \hat m^2_{(\pm)}-\hat \triangle
 ]D_{(\pm)}(x,y)&=&\delta^3(x-y),\\\label{2-14}
\hat m^2_{(\pm)}&=& 14 (\beta\pm 1)\langle \tau_{(0)}\rangle \mp
\langle \tau_{(1)}\rangle.
 \eea
The reduced Hamiltonian
 function (\ref{hf1}) after its decomposition (\ref{pt1})  takes the form of the current-current interaction
\bea\label{hf2} E_\vh&=&2\int d^3x\sqrt{{T^0_0}}=
2V_0\sqrt{\langle T_0\rangle}\left[1+
  \frac{\langle T_2\rangle}{2\langle T_0\rangle}-
  \frac{\langle T^2_1\rangle}{8\langle
  T_0\rangle^2}+...\right]=\\\nonumber
 &=&2V_0\sqrt{\langle \tau_{(0)}\rangle}\left[1+\frac{1}{28\beta\langle \tau_{(0)}\rangle}
 \left\langle\left(\overline{J_{(+)}}D_{(+)}\cdot \overline{J_{(+)}}
 +\overline{J_{(-)}}D_{(-)}\cdot \overline{J_{(-)}}\right)\right\rangle\right].
 \eea
   In the case of point mass distribution in a finite volume $V_0$ with the zero pressure
  and  the  density
  $\overline{\tau_{(1)}}=\overline{\tau_{(2)}}
/6\equiv \sum_J M_J\left[\delta^3(x-y_J)-\frac{1}{V_0}\right]$,
 solutions   (\ref{2-17}),  (\ref{2-18}) take
 the very significant form:
 \bea\label{2-21}
  \mu(x)&=\sum_{J}
  \frac{r_{gJ}}{4r_{J}}\left[{\gamma_1}e^{-m_{(+)}(z)
 r_{J}}+ (1-\gamma_1)\cos{m_{(-)}(z)
 r_{J}}\right],\\\label{2-22}
 \nu(x)&=\sum_{J}
 \frac{2r_{gJ}}{r_{J}}\left[(1-\gamma_2)e^{-m_{(+)}(z)
 r_{J}}+ {\gamma_2}\cos{m_{(-)}(z)
 r_{J}}]\right],
 \eea
 where
 $$
  {\gamma_1}=\frac{1+7\beta}{14\beta},~~~
 {\gamma_2}=\frac{(1-\beta)(7\beta-1)}{16\beta},~~
 r_{gJ}=\frac{3M_J}{4\pi\vh^2},~~
 r_{J}=|x-y_J|.
 $$
 The minimal surface (\ref{dgauge})
 gives the shift of the coordinate
  origin in the process of evolution $\partial_i{\cal
 N}^i=6\overline{\mu}'$:
 \be \label{2-23}
 {\cal N}^i=\sum_{J}\frac{3r_{gI}'}{4}
 \frac{(x-y_J)^i}{|x-y_J|}\left[{\gamma_1}e^{-m_{(+)}(z)
 r_{J}}+(1-\gamma_1) \cos{m_{(-)}(z)
 r_{J}}\right].
  \ee
In the infinite volume limit $\langle \tau_{(n)}\rangle=0$ these
 solutions take the standard Newtonian form:
 $\mu=D\cdot \tau_{(0)}$, $\nu=D\cdot [14\tau_{(0)}-\tau_{(1)}]$,
 ${\cal N}^i=0$
 (where $\hat \triangle D(x)=-\delta^3(x)$).
 However, the isotropic version of Schwarzschild solutions
  $e^\mu=1+\frac{r_g}{4r};~~\frac{e^{7\mu}}{1+\nu}
  =1-\frac{r_g}{4r}$ of equations
  $\hat \triangle e^\mu=0, ~\hat \triangle \frac{e^{7\mu}}{1+\nu}=0$
  obtained in the infinite volume do not  satisfy  Eqs.
  (\ref{14ec}), (\ref{4f4}) in the presence of cosmological
  background.
  Therefore, the exact resolution  of these equations does not commute with the infinite-volume
 limit.
 The d-invariant analog of the Schwarzschild solution takes the
 form
\be\label{33}
~~~~~~~~~~ds^2=a^2(\zeta)\left[\frac{e^{12\mu}}{1+\nu}d\zeta^2-e^{4\mu}(dx^i+{\cal
N}^id\zeta)^2\right], \ee where $\mu,\nu, {\cal N}^i$ are given by
Eqs. (\ref{2-21}) -- (\ref{2-23}) and  $\langle\mu\rangle=0$,
$\langle\nu\rangle=0$.

 The reduced action (\ref{2ha2}) determines
   evolution of fields directly in terms of the cosmological scale factor $a=\vh/\vh_0$
  connected with the red shift parameter
   $z$ by the relation
  $\vh=\vh_0/(1+z)$. Therefore, the d-invariant perturbation
  theory for the reduced Hamiltonian function  (\ref{hf2}) converts
  into the d-invariant  cosmological perturbation
  theory, if all spatial ``averagings`` $\langle \tau_{(n)}\rangle$ in (\ref{hf2})
  are
    sums $\langle \tau_{(n)}\rangle=\rho_{(n)}(\vh)+t^{\rm f}_{(n)}$
   of small field parts  $t^{\rm f}_{(n)}$ associated with
    the Standard Model (SM)
   and the tremendous ($\sim 10^{79}$~GeV) cosmological background
   $\rho_{(n)}(\vh) \equiv H_0^2\vh_0^2\Omega_{(n)}$, where
   $H_0$ is the present day value of the Hubble parameter and
   $
 \Omega_{(n)} \equiv \sum_{I} {I}^n
 \Omega_I (1+z)^{2-I/2}
 $ 
  here
 $\Omega_I
$ is the partial cosmological densities normalized by condition
$\Omega_{(0)}=\sum_I\Omega_I=1$ (recall that  the index
 $I$ runs a set of values
   I=0 (stiff), 4 (radiation), 6 (mass), 8 (curvature), 12 ($\Lambda$-term)
in the correspondence with the type of physical contributions).
 In the presence of the tremendous  cosmological background one
 can apply the
 Einstein correspondence principle \cite{pp} as
 the low-energy decomposition
  of ``reduced''  action (\ref{2ha2}) $d\vh 2\sqrt{\langle \tau_{(0)}\rangle}= d\vh
 2\sqrt{\rho_{(0)}+t^{\rm f}_{(0)}}
 =
 d\vh
 \left[2\sqrt{\rho_{(0)}}+
 t^{\rm f}_{(0)}/{\sqrt{\rho_{(0)}}}\right]+...$
 over
 field density $t^{\rm f}_{(0)}$. The first  term of these sum
  $S^{(\pm)}|_{P_\vh=\pm E_\vh
 }= S^{(\pm)}_{\rm cosmic}+S^{(\pm)}_{\rm
 field}+\ldots$ is  the reduced  cosmological action
 $S^{(\pm)}_{\rm cosmic}= \mp
 2V_0\int\limits_{\vh_I}^{\vh_0}\!
 d\vh\!\sqrt{\rho_{(0)}(\vh)}$; whereas the second one is
  the standard field action of GR and SM
 \be\label{12h5} S^{(\pm)}_{\rm field}=
 \int\limits_{\eta_I}^{\eta_0} d\eta\int d^3x
 \left[\sum\limits_{ F}P_{ F}\partial_\eta F
 +\bar{{\cal C}}-t^{\rm f}_{(0)} \right]
 \ee
 in the  space determined by the interval
 \be\label{d2}
 ds^2=d\eta^2-[e_{(a)i}(dx^i+{\cal N}^id\eta]^2;
 ~~\partial_ie^i_{(a)}=0,~~\partial_i{\cal N}^i=0
 \ee
 with  conformal time
 $d\eta=d\vh/\rho_{(0)}^{1/2}$
 and running masses
 $m(\eta)=a(\eta)m_0$ \cite{ps1}.

 We see that
  the  correspondence principle leads to the theory (\ref{12h5}),
  where the conformal  variables and coordinates are
    identified  with observable ones
    and the cosmic evolution with the evolution of masses.
    The best fit to the data  included
  high-redshift Type Ia supernovae \cite{SN} 
 requires a cosmological constant $\Omega_{\Lambda}=0.7$,
$\Omega_{\rm Cold Dark Matter}=0.3$ in the case of the Friedmann
``absolute quantities`` of standard cosmology, whereas for
``conformal
 quantities'' of the d-invariant approach
 these data are consistent with  the dominance of the stiff state
of free scalar field  $[z +1]^{-1}|_{({\rm stiff})}(\eta)
 =\sqrt{1+2H_0(\eta-\eta_0)}$: $\Omega_{\rm Stiff}=0.85\pm 0.15$,
$\Omega_{\rm Cold Dark Matter}=0.15\pm 0.10$ \cite{039}.

 The d-invariant reduction allows
 us to consider on equal footing the quantum field theories  of both
 particles and universes
 in the framework of
 perturbation
 theory $t^{\rm f}_{(0)}=t_{\rm free}+t_{\rm int}$ extracting the
 free part $t_{\rm free}=\sum_p E_p a_p^+a_p^- $ with
 one-particle
 energies $E_p=\sqrt{p^2+m^2(\eta)}$ \cite{ps1,114:a}  and defining  Quantum Cosmology
  as the
 primary quantization of the energy constraint
 $P^2_{\vh}-E^2_\vh=0$:
 $\partial^2_\vh\Psi+E_\vh^2\Psi=0$ and the secondary one:
 $\Psi=[A^++A^-]/({\sqrt{2E_\vh}})$  with the reduced energy
 $E_\vh=2V_0\sqrt{\rho_{(0)}(\vh)}$, when   stable vacua
 of both universes (u) $B^-|0>_{u}=0$
  and particles (p)
 $b_p^-|0>_{p}=0$ are determined by
   Bogoliubov's
 transformations of ``universes'' $A^+=\alpha_u
 B^+\!+\!\beta_u^*B^-$ and particles:
 $a_p^+=\alpha_p b_p^+\!+\!\beta_p^*b_p^-$.
 The ``vacuum'' postulate
leads to positive arrow of the time interval (\ref{13c})
$\eta_{\pm}\geq 0$ and its absolute beginning \cite{pp}. In the
case  of the stiff state: $E_\vh=Q/\vh$, there is an exact
solution of  Bogoliubov's equation of a number of ``universes''
\cite{origin}
 \be\label{cu}
 {}_{\rm u}\!\!<0|A^+A^-|0>_{\rm u}=\frac{1}{4Q^2-1}
 \sin^2\left[\sqrt{Q^2-\frac{1}{4}}~~\ln\frac{\vh_0}{\vh_I}\right]\not
=0, \ee
 where the Planck mass $\vh_0=\vh_I\sqrt{1+2H_I\eta_0}$
 belongs to the present-day data $\eta=\eta_0$ and
 $\vh_I,H_I=\vh'_I/\vh_I=Q/(2V_0\vh_I^2)$ are the initial data.

\begin{figure}
  \centering
  \includegraphics{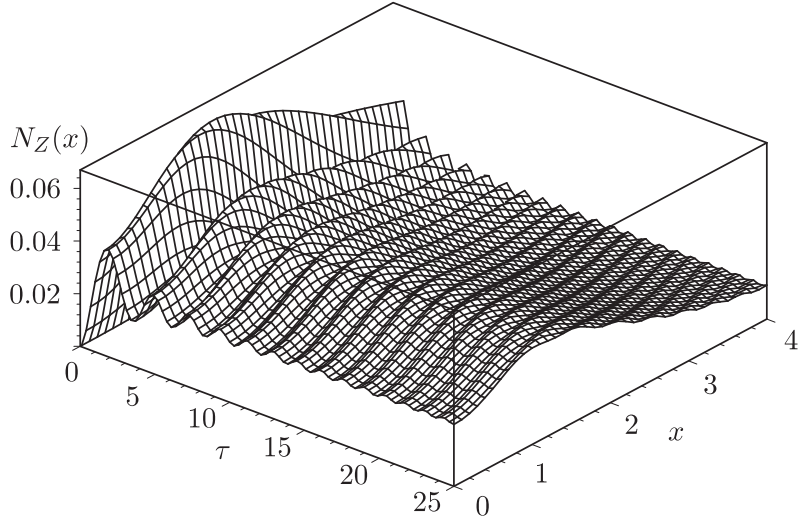}
  \caption{Longitudinal ($N_Z(x)$) components
of the boson distribution versus the dimensionless time $\tau=
2\eta H_I$ and the dimensionless momentum $x = q/M_I$ at the
initial data $M_I = H_I$ ($\gamma_v = 1$).}
\end{figure}

 These initial data $\vh_I$ and $H_I$ are determined by
 parameters of matter cosmologically created from the Bogoliubov
 vacuum  in beginning of a universe $\eta\simeq 0$.
 In the Standard
 Model (SM),
   W-, Z- vector bosons have maximal probability of this
 cosmological creation
 due to their mass singularity~\cite{114:a}. The uncertainty principle
  $\triangle E\cdot\triangle \eta \geq 1$ (where $\triangle E=2M_I,\triangle \eta=1/2H_I$)
  shows us that at
   the moment of creation of  vector bosons their  Compton
   lengths
 defined by its inverse mass
 $M^{-1}_{\rm I}=(a_{\rm I} M_{\rm W})^{-1}$ are close to the
 universe horizon defined in the
 stiff state as
 $H_{\rm I}^{-1}=a^2_{\rm I} (H_{0})^{-1}$.
 Equating these quantities $M_{\rm I}=H_{\rm I}$
 one can estimate the initial data of the scale factor
 $a_{\rm I}^2=(H_0/M_{\rm W})^{2/3}=10^{29}$ and the Hubble parameter
 $H_{\rm I}=10^{29}H_0\sim 1 {\rm mm}^{-1}\sim 3 K$.
 Just at this moment there is  an effect of intensive
  cosmological creation of the vector bosons described in \cite{114:a};
 in particular, the distribution functions of the longitudinal   vector bosons
 (see Fig. 1)
demonstrate us  large contribution of relativistic momenta. Their
temperature $T_c$  can be estimated from the equation in the
kinetic theory for the time of establishment of this temperature $
\eta^{-1}_{relaxation}\sim n(T_c)\times \sigma \sim H $, where
$n(T_c)\sim T_c^3$ and $\sigma \sim 1/M^2$ is the cross-section.
This kinetic equation and values of the initial data $M_{\rm I} =
H_{\rm I}$ give the temperature of relativistic bosons $
 T_c\sim (M_{\rm I}^2H_{\rm I})^{1/3}=(M_0^2H_0)^{1/3}\sim 3 K
$ as a conserved number of cosmic evolution compatible with the
Supernova data \cite{039}.
 We can see that
this  value surprisingly close to the observed temperature of the
CMB radiation
 $ T_c=T_{\rm CMB}= 2.73~{\rm K}$. The primordial mesons before
 their decays polarize the Dirac fermion vacuum and give the
 baryon asymmetry frozen by the CP -- violation,
 so that $n_b/n_\gamma \sim X_{CP} \sim 10^{-9}$,
 $\Omega_b \sim \alpha_{\rm qed}/\sin^2\theta_{\rm Weinberg}\sim
 0.03$, and $\Omega_R\sim 10^{-5}\div 10^{-4}$~\cite{114:a}.

\section{Two approaches to descriptions of CMBR fluctuation}

 Now one can
  compare the d-invariant perturbation theory  (\ref{33}) with  the
 standard cosmological perturbation theory
 \cite{lif}
  $ds^2=a^2(\eta)[(1+\overline{\Phi})d\eta^2-(1-\overline{\Psi})
 (dx^i+N^id\eta)^2]$, $\overline{\Phi}=\nu-6\mu$ ,
 $\overline{\Psi}=2\mu$, $N^i=0$,
   where the zero-Fourier harmonics
 of the spatial determinant is taking into account twofold that
 is an obstruction to the Dirac Hamiltonian method. The d-invariant perturbation theory
 shows us that, if this double counting is removed, then equations
 of  scalar potential $\mu$ and $\nu$ (see Eqs. (\ref{14ec}),
  (\ref{4f4}))
 do not contain time derivatives that are responsible for
the CMB ``primordial power spectrum'' in the inflationary model
\cite{bard}. However, the d-invariant version of the Dirac
Hamiltonian approach to GR gives us another possibility to explain
the CMBR ``spectrum'' and other topical problems of cosmology by
cosmological creation of the vector bosons considered above. The
equations describing the longitudinal vector bosons
 in SM, in this case, are close to
 the equations that follows from
 the Lifshitz perturbation theory and are  used, in  the inflationary model, for
 description of the ``power primordial spectrum'' of the CMB radiation.

 The next differences are a nonzero shift vector and the spatial oscillations of
 the scalar potentials determined by $\hat m^2_{(-)}$
 In the d-invariant version of cosmology \cite{039}, the
  SN data dominance of stiff state $\Omega_{\rm Stiff}\sim 1$ determines the parameter
  of spatial oscillations $\hat m^2_{(-)}=\frac{6}{7}H_0^2[\Omega_{\rm R}(z+1)^2+\frac{9}{2}\Omega_{\rm
  Mass}(z+1)]$. The values of red shift in the recombination
  epoch $z_r\sim 1100$ and the clusterization parameter \cite{kl}
 $
 r_{\rm clusterization}=\frac{\pi}{\hat m_{(-)} }\sim \frac{\pi}{
 H_0\Omega_R^{1/2} (1+z_r)} \sim 130\, {\rm Mpc}
 $
  recently
 discovered in the researches of large scale periodicity in redshift
 distribution
 lead to reasonable value of the radiation-type density
  $10^{-4}<\Omega_R\sim 3\cdot 10^{-3}<5\cdot 10^{-2}$ at the time of this
  epoch.


\section{\label{a} Conclusions}

   The diffeomorphism-invariant
Hamiltonian approach to GR with an exact solution of energy
constraint in a finite volume was applied
 to formulate the corresponding cosmological perturbation theory.
The  Hamiltonian  cosmological perturbation theory
 would be considered
   as the  foundation of
    the standard cosmological perturbation theory \cite{lif,bard},
     if it did not contain the double counting of the scale factor
     as an obstruction to the Dirac Hamiltonian method.
 Avoiding this  double counting
 we obtained new Hamiltonian  equations. These  equations do not contain
 the time derivatives
 that are responsible for the ``primordial power spectrum'' in
 the inflationary model \cite{bard}.
However, the Hamiltonian approach to GR gives  us
 another  possibility  to explain this ``spectrum'' and
  other topical problems of cosmology by
 the cosmological creation of the primordial W-, Z- bosons
  from vacuum due to their mass singularity, when
 their Compton length coincides with the universe horizon.

{\bf Acknowledgement}\\
 The authors are grateful to  A.~Gusev, A. Efremov, E. Mychelkin,
  E. Kuraev,
 L.~Lipatov,  V. Priezzhev, G. Vereshkov,  and  S. Vinitsky
 for fruitful discussions.


\begin{thebibliography}{999}


\bibitem{lif}
E.~Lifshitz, {Zh. Exp. Teor. Fiz.}  {\bf 16}, 587 (1946); J.
M.~Bardeen, Phys. Rev. {\bf D 22}, 1882 (1980); H.~Kodama,
M.~Sasaki, Prog. Theor. Phys., {N \bf 78}, 1 (1984).
\bibitem{bard}
 V. F. Mukhanov, H.A. Feldman and R.H. Brandenberger, Phys. Rep. {\bf 215}, 206 (1992).

 \bibitem{prd}  A.M.~Khvedelidze, V.V.~Papoyan, V.N.~Pervushin, Phys. Rev.  {\bf D
 51}, 5654 (1995).

\bibitem{ps1}
V.N.~Pervushin and V.I.~Smirichinski, J. Phys. A: Math. Gen. {\bf
32}, 6191 (1999).

\bibitem{pp}
M.~Pawlowski, V.N.~Pervushin, Int. J. Mod. Phys. {\bf 16}, 1715
(2001); [hep-th/0006116].
\bibitem{bpp}
B.M. Barbashov, V.N. Pervushin, and D.V. Proskurin, {Theoretical
and Mathematical Physics} {\bf 132}, 1045 (2002).

\bibitem{dir}
 P.A.M.~Dirac, Proc. Roy. Soc. {\bf A 246}, 333 (1958);
  Phys. Rev. {\bf 114}, 924 (1959).

 \bibitem{ADM}
R.~Arnowitt, S.~Deser and C.W.~Misner, Phys. Rev. {\bf 117}, 1595
(1960).




\bibitem{vlad}
A.L.~Zelmanov, 
 Dokl. AN USSR {\bf 209}, 822 (1973).

\bibitem{L}
A. Lichnerowicz, Journ. Math. Pures and Appl. {\bf B 37}, 23
(1944).
\bibitem{Y}
J.W.~York. (Jr.), Phys. Rev. Lett. {\bf 26}, 1658 (1971).
\bibitem{fock29} V.A.~Fock, Zs. f. Phys. {\bf 57}, 261 (1929).


\bibitem{SN}
A. G. Riess {\it et al.}, Astron. J. {\bf 116}, 1009 (1998) ; S.
Perlmutter {\it et al.}, Astrophys. J. {\bf 517}, 565 (1999); A.
G. Riess {\it et al.}, Astrophys. J. {\bf 560}, 49 (2001).

\bibitem{039}
   D.Behnke, D.Blaschke, V.P.,  D.Proskurin,
Phys. Lett. {\bf B 530}, 20 (2002);[gr-qc/0102039].   D. Behnke,
PhD Thesis, Rostock Report MPG-VT-UR 248/04 (2004)


\bibitem{114:a}
D. B. Blaschke, S. I. Vinitsky, A. A. Gusev, V .N. Pervushin, and
D. V. Proskurin, Physics of Atomic Nuclei {\bf 67},
 1050 (2004); [hep-ph/0504225].
\bibitem{origin} V. Pervushin, V. Zinchuk, Talk in  XI St.Petersburg School on Theoretical
Physics, qr-qc/0504123.
\bibitem{kl} V.I. Klyatskin, {\it Stochastic Equations}, Moscow,
 Fizmatlit, 2001 (in Russian).
\bibitem{a1}
W. J. Cocke and W. G. Tifft,  ApJ. {\bf 368}, 383 (1991); K.
Bajan, P. Flin,  W. God{\l}owski, V. Pervushin, and A. Zorin,
Spacetime \& Substance {\bf 4},  225 (2003).
\end{thebibliography}
\end{document}